# A Three-Dimensional Quantum Simulation of Silicon Nanowire Transistors with the Effective-Mass Approximation

Jing Wang[*], Eric Polizzi[**] and Mark Lundstrom[*]

[*]School of Electrical and Computer Engineering, Purdue University, West Lafayette, Indiana 47907, USA

[**]Department of Computer Sciences, Purdue University, West Lafayette, Indiana 47907, USA

## ABSTRACT

The silicon nanowire transistor (SNWT) is a promising device structure for future integrated circuits, and simulations will be important for understanding its device physics and assessing its ultimate performance limits. In this work, we present a three-dimensional quantum mechanical simulation approach to treat various SNWTs within the effective-mass approximation. We begin by assuming ballistic transport, which gives the upper performance limit of the devices. The use of a mode space approach (either coupled or uncoupled) produces high computational efficiency that makes our 3D quantum simulator practical for extensive device simulation and design. Scattering in SNWTs is then treated by a simple model that uses so-called Büttiker probes, which was previously used in metal-oxide-semiconductor field effect transistor (MOSFET) simulations. Using this simple approach, the effects of scattering on both internal device characteristics and terminal currents can be examined, which enables our simulator to be used for the exploration of realistic performance limits of SNWTs.

# I. INTRODUCTION

As the channel lengths of metal-oxide-semiconductor field effect transistors (MOSFETs) scale into the nanometer regime, short channel effects[1] become more and more significant. Consequently, effective gate control is required for a nanoscale MOSFET to achieve good device performance. For this reason, silicon nanowires, which allow multi-gate or gate-all-around transistors, are being explored.[2-7] In Ref. [2], the authors reported a parallel wire channel transistor, whose channel can be viewed as a wire with a triangular cross-section. In Refs. [3-6], wires with rectangular cross-sections were used to fabricate different types of tri-gate/gate-all-around FETs. At the same time, cylindrical Si nanowires with diameters as small as 5nm have also been synthesized by the chemical vapor deposition technology.[7] These recent experiments have shed light on the potential applications of silicon nanowire transistors in future electronics.

To deeply understand device physics of silicon nanowire transistors (SNWTs) and to assess their ultimate performance limits, simulation work is necessary and important. In contrast to a planar MOSFET, which has a uniform charge and potential profile in the transverse direction (normal to both the gate and the source-to-drain direction), an SNWT has a three-dimensional (3D) distribution of electron density and electrostatic potential. As a result, a 3D simulator is required for the simulation of SNWTs. In this paper, we propose a 3D self-consistent quantum simulation of SNWTs based on the effective-mass approximation (whose validity in the nanoscale device simulation has been established in Ref. [8]). The calculation involves a self-consistent solution of a 3D Poisson equation and a 3D Schrödinger equation with open boundary conditions. Using the finite element method (FEM), we solve the 3D Poisson equation to obtain the electrostatic potential. At the same time, we solve the 3D Schrödinger by a (coupled/uncoupled) mode space approach.[9-12] which provides both computational efficiency and high accuracy as compared with direct real space calculations. Since the (coupled/uncoupled) mode space approach treats quantum confinement and transport separately, the procedure of the calculation is as follows:

Step 1: Solve the 3D Poisson equation for the electrostatic potential;

Step 2: Solve a 2D Schrödinger equation with a closed boundary condition at each slice (cross-section) of the nanowire transistor (see FIG. 1) to obtain the electron subbands (along the nanowire) and eigenfunctions;

Step 3: Solve (coupled/uncoupled) 1D transport equations by the nonequilibrium Green's function (NEGF) approach[13-15] for the electron charge density;

Step 4: Go back to Step 1 to calculate the electrostatic potential. If it converges, then calculate the electron current by the NEGF approach (as in Step 3) and output the results. Otherwise continue Step 2 and 3.

Different transport models (in Step 3) are implemented into our simulator. In this paper, we will discuss both ballistic NEGF model, which gives the upper performance limit of SNWTs, and a dissipative NEGF model with a simple treatment of scattering with the Büttiker probes,[14,16,17] which offers an efficient way to capture scattering in the quantum mechanical framework.

A rigorous treatment of scattering and a detailed calculation of bandstructures are very important to understand physics in Si nanowires in detail. However, the huge computational cost involved in such a rigorous model can prevent it from being used for extensive device simulation and design. As we will show later, the use of the effective-mass approximation and the simple treatment of scattering with the Büttiker probes greatly reduces the computational complexity while still capturing the essential device physics of SNWTs (i.e., 3D electrostatics, quantum confinement, source-to-tunneling and scattering, etc), so the method we discuss in this paper can be used as a practical 3D quantum approach for device study and design of SNWTs.[18] This paper is divided into the following sections: Sec. II describes our methodology for ballistic SNWTs and provides the basic equations, Sec. III discusses the simulation results for ballistic SNWTs with arbitrary cross-sections (e.g., triangular, rectangular and cylindrical), Sec. IV introduces the Büttiker probes for the treatment of scattering and shows relevant results, and Sec. V summarizes key findings.

# II. THEORY FOR BALLISTIC SILICON NANOWIRE TRANSISTORS

Figure 1 shows a schematic structure of the Si nanowire transistors simulated in this work. This intrinsic device structure is connected to two infinite reservoirs, the source and drain (not shown), so the source/drain (S/D) extension regions are terminated using open boundary conditions. As shown in FIG. 1 (b), a uniform grid with a grid spacing of $a$ is used along the channel (x) direction. In the y-z plane (the cross-section of the SNWT), a 2D finite element mesh with triangular elements is generated by Easymesh-1.4,[19] which allows us to treat nanowires with arbitrary cross-sections (e.g., triangular, rectangular and cylindrical, as shown in FIG. 1 (c)). By doing this, a 3D finite element mesh with prism elements is constructed. When solving the Poisson equation, the 3D Laplacian is directly discretized by the FEM approach. The obtained linear system is solved using a Preconditioned Conjugate Gradient method with Incomplete Cholesky factorization. More details about the numerical techniques can be found in Ref. [9].

As mentioned earlier, we solve the 3D Schrödinger equation by the mode space approach,[9-11] which is based on an expansion of the active device Hamiltonian in the subband eigenfunction space. As a result, we need to solve a 2D Schrödinger equation by the FEM at each slice of the SNWT to obtain the subband eigenenergy levels and eigenfunctions (modes). After that, the original 3D device Hamiltonian is transformed into a 1D Hamiltonian in the x direction, which can be used to calculate electron density and current within the NEGF formalism. In this section, we will first give an overview of the coupled mode space (CMS) approach for the SNWT simulation (part A), which is mathematically equivalent to a direct real space solution if adequate modes are included (to be discussed later).[9,10] Then we will introduce the uncoupled mode space (UMS) approach (part B) and a fast uncoupled mode space (FUMS) approach (part C), which are a simplification of the CMS approach to provide high computational efficiency. The simulation results (in Sec. III) illustrate that the UMS and FUMS approaches show excellent agreement with the CMS approach for the SNWT simulation.

## A. The coupled mode space (CMS) approach

In this part of the work, we will briefly review the Coupled Mode Space (CMS) approach and list basic equations for our particular case of interest.

In the 3D domain, the full stationery Schrödinger equation is given by

$$H_{3D}\Psi(x,y,z) = E\Psi(x,y,z),$$ (1)

where $H_{3D}$ is the 3D device Hamiltonian. Assuming an ellipsoidal parabolic energy band with a diagonal effective-mass tensor (for the case that the effective-mass tensor includes non-zero off-diagonal elements, please refer to Ref. [12]), $H_{3D}$ is defined as

$$H_{3D} = -\frac{\hbar^2}{2m_x^*(y,z)}\frac{\partial^2}{\partial x^2} - \frac{\hbar^2}{2}\frac{\partial}{\partial y}\left(\frac{1}{m_y^*(y,z)}\frac{\partial}{\partial y}\right) - \frac{\hbar^2}{2}\frac{\partial}{\partial z}\left(\frac{1}{m_z^*(y,z)}\frac{\partial}{\partial z}\right) + U(x,y,z),$$ (2)

here $m_x^*$, $m_y^*$ and $m_z^*$ are the electron effective mass in the x, y, and z directions, respectively, and $U(x,y,z)$ is the electron conduction band-edge profile in the active device. We note that the effective mass varies in the y and z directions due to the transition between the Si body and the SiO$_2$ layer. Now let us expand the 3D electron wavefunction in the subband eigenfunction space,

$$\Psi(x,y,z) = \sum_n \varphi^n(x) \cdot \xi^n(y,z;x),$$ (3)

where $\xi^n(y,z;x = x_0)$ is the $n^{th}$ eigenfunction of the following 2D Schrödinger equation at the slice ($x = x_0$) of the SNWT,

$$\left[-\frac{\hbar^2}{2}\frac{\partial}{\partial y}\left(\frac{1}{m_y^*(y,z)}\frac{\partial}{\partial y}\right) - \frac{\hbar^2}{2}\frac{\partial}{\partial z}\left(\frac{1}{m_z^*(y,z)}\frac{\partial}{\partial z}\right) + U(x_0,y,z)\right]\xi^n(y,z;x_0) = E_{sub}^n(x_0)\xi^n(y,z;x_0),$$ (4)

here $E_{sub}^n(x_0)$ is the $n^{th}$ subband energy level at $x = x_0$. According to the property of eigenfunctions, $\xi^n(y,z;x)$ satisfies the following equation for any $x$,

$$\oint_{y,z} \xi^m(y,z;x)\xi^n(y,z;x)dydz = \delta_{m,n},$$ (5)

where $\delta_{m,n}$ is the Kronecker delta function.

Inserting Eqs. (2) and (3) into Eq. (1) and using the relation described by Eq. (4), we obtain

$$-\frac{\hbar^2}{2m_x^*(y,z)}\frac{\partial^2}{\partial x^2}\left(\sum_n \varphi^n(x)\cdot\xi^n(y,z;x)\right) + \sum_n \varphi^n(x)\cdot E_{sub}^n(x)\xi^n(y,z;x) = E\sum_n \varphi^n(x)\cdot\xi^n(y,z;x).$$ (6)

Now we multiply by $\xi^m(y,z;x)$ on both sides and do an integral within the y-z plane. According to Eq. (5), we obtain the following 1D coupled Schrödinger equation

$$-\frac{\hbar^2}{2}\left(\sum_n a_{mn}(x)\right)\frac{\partial^2}{\partial x^2}\varphi^m(x) - \frac{\hbar^2}{2}\sum_n c_{mn}(x)\varphi^n(x) - \hbar^2\sum_n b_{mn}(x)\frac{\partial}{\partial x}\varphi^n(x) + E_{sub}^m(x)\varphi^m(x) = E\varphi^m(x) \text{ , (7)}$$

where

$$a_{mn}(x) = \oint_{y,z}\frac{1}{m_x^*(y,z)}\xi^m(y,z;x)\xi^n(y,z;x)dydz \text{ ,} \tag{8a}$$

$$b_{mn}(x) = \oint_{y,z}\frac{1}{m_x^*(y,z)}\xi^m(y,z;x)\frac{\partial}{\partial x}\xi^n(y,z;x)dydz \text{ ,} \tag{8b}$$

and

$$c_{mn}(x) = \oint_{y,z}\frac{1}{m_x^*(y,z)}\xi^m(y,z;x)\frac{\partial^2}{\partial x^2}\xi^n(y,z;x)dydz \text{ .} \tag{8c}$$

Eq. (7) is the basic equation for the CMS approach. In our simulation, since the electron wavefunction is mainly located in the silicon, we can neglect $a_{mn}$ if $m \neq n$ ($a_{mm} \gg a_{mn}$)[10] and simplify Eq. (7) as

$$-\frac{\hbar^2}{2}a_{mm}(x)\frac{\partial^2}{\partial x^2}\varphi^m(x) - \frac{\hbar^2}{2}\sum_n c_{mn}(x)\varphi^n(x) - \hbar^2\sum_n b_{mn}(x)\frac{\partial}{\partial x}\varphi^n(x) + E_{sub}^m(x)\varphi^m(x) = E\varphi^m(x) \text{ . (9)}$$

From the derivation above, it is clear that the CMS formalism (Eqs. (7) and (8)) is mathematically equivalent to the real space calculation if all the modes (i.e., $m,n = 1,...,N_{YZ}$, where $N_{YZ}$ is the number of nodes in the y-z plane) are included. In practice, due to strong quantum confinement in SNWTs, usually only a few of the lowest subbands (i.e., $m,n = 1,...,M$, $M \ll N_{YZ}$) are occupied and need to be included in the calculation (which means that if we increase the mode number, $M$, the device characteristics such as the electron density profile and terminal currents will not change any more). Thus, with the first $M$ subbands considered (i.e., $m,n=1,...,M$), Eq. (9) represents an equation group that contains $M$ equations, each representing a selected mode. We can write down these $M$ equations in a matrix format

$$H \begin{bmatrix} \varphi^1(x) \\ \varphi^2(x) \\ ... \\ ... \\ \varphi^M(x) \end{bmatrix} = \begin{bmatrix} h_{11} & h_{12} & h_{13} & ... & h_{1M} \\ h_{21} & h_{22} & h_{23} & ... & h_{2M} \\ ... & ... & ... & ... & ... \\ ... & ... & ... & ... & ... \\ h_{M1} & h_{M2} & h_{M3} & ... & h_{MM} \end{bmatrix} \begin{bmatrix} \varphi^1(x) \\ \varphi^2(x) \\ ... \\ ... \\ \varphi^M(x) \end{bmatrix} = E \begin{bmatrix} \varphi^1(x) \\ \varphi^2(x) \\ ... \\ ... \\ \varphi^M(x) \end{bmatrix}, \tag{10}$$

where

$$h_{mn} = \delta_{m,n} \left[ -\frac{\hbar^2}{2} a_{mm}(x) \frac{\partial^2}{\partial x^2} + E_{sub}^m(x) \right] - \frac{\hbar^2}{2} c_{mn}(x) - \hbar^2 b_{mn}(x) \frac{\partial}{\partial x}, \, (m,n=1,2,...,M). \tag{11}$$

By using the coupled mode space approach, the size of the device Hamiltonian, $H$, has been reduced to $M \cdot N_X \times M \cdot N_X$ ( $N_X$ is the number of nodes in the x direction, and the mode number $M$ we need is normally less than 5 for the SNWT structures we simulate), which is much smaller than that in the real space representation, $N_{YZ} \cdot N_X \times N_{YZ} \cdot N_X$ ( $N_{YZ}$ is ~1,000 for the device structures simulated in this work).

After the device Hamiltonian $H$ is obtained, we can calculate the electron density and current using the NEGF approach. The NEGF approach, a widely used method for the simulation of nanoscale electronic devices, has been discussed in Refs. [13, 14]. Here we list the relevant equations for our particular case.

The retarded Green's function of the active device is defined as[14]

$$G(E) = \left[ ES - H - \Sigma_S(E) - \Sigma_1(E) - \Sigma_2(E) \right]^{-1}, \tag{12}$$

where the device Hamiltonian, $H$, is defined by Eq. (10), $\Sigma_S$ is the self-energy that accounts for the scattering inside the device (in the ballistic limit, it is equal to zero), $\Sigma_1$ ($\Sigma_2$) is the self-energy caused by the coupling between the device and the source (drain) reservoir. If we discretize the equations by the 1D (in the x direction) Finite Difference Method (FDM), the matrix $S$ in Eq. (12) is equal to an $M \cdot N_X \times M \cdot N_X$ identity matrix. The self-energies, $\Sigma_1$ and $\Sigma_2$, are defined as[14]

$$\Sigma_1[p,q] = -t_{m,1} \exp\left( jk_{m,1}a \right) \delta_{p,(m-1)N_X+1} \delta_{q,(m-1)N_X+1}, \, ( j=\sqrt{-1} \text{ ) (FDM)}, \tag{13}$$

$$\Sigma_2[p,q] = -t_{m,N_X} \exp\left( jk_{m,N_X}a \right) \delta_{p,mN_X} \delta_{q,mN_X}, \, (m=1,2,..., M \text{ and } p,q=1,2,...,MN_X), \text{ (FDM)}, \tag{14}$$

where $t_{m,1} = \left(\hbar^2/2a^2\right) a_{mm}(x)\big|_{x=0}$ and $t_{m,N_X} = \left(\hbar^2/2a^2\right) a_{mm}(x)\big|_{x=(N_X-1)a}$ ($a_{mm}(x)$ is defined by Eq.(8a)), and $k_{m,1}$ ($k_{m,N_X}$) is determined by $E = E_{sub}^m(0) + 2t_{m,1}\left(1 - \cos k_{m,1}a\right)$ ($E = E_{sub}^m\left((N_X-1)a\right) + 2t_{m,N_X}\left(1 - \cos k_{m,N_X}a\right)$).

If we discretize the equations by the 1D (in the x direction) Finite Element Method (FEM), the matrix $S$ in Eq. (12) becomes a $M \cdot N_X \times M \cdot N_X$ block diagonal matrix

$$S = \begin{bmatrix} S_0 & 0 & \cdots & \cdots & 0 \\ 0 & S_0 & 0 & \ddots & \vdots \\ \vdots & 0 & \ddots & \ddots & \vdots \\ \vdots & \ddots & \ddots & \ddots & 0 \\ 0 & \cdots & \cdots & 0 & S_0 \end{bmatrix} \text{ (FEM)}, \tag{14}$$

where $S_0$ is a $N_X \times N_X$ matrix[15]

$$S_0 = \begin{bmatrix} a/3 & a/6 & 0 & \cdots & \cdots & 0 \\ a/6 & 2a/3 & a/6 & \ddots & \ddots & \vdots \\ 0 & a/6 & 2a/3 & \ddots & \ddots & \vdots \\ \vdots & \ddots & \ddots & \ddots & \ddots & 0 \\ \vdots & \ddots & \ddots & \ddots & 2a/3 & a/6 \\ 0 & \cdots & \cdots & 0 & a/6 & a/3 \end{bmatrix} \text{ (FEM)}. \tag{16}$$

The self-energies, $\Sigma_1$ and $\Sigma_2$, are defined as[10,15]

$$\Sigma_1[p,q] = -jk_{m,1}at_{m,1}\delta_{p,(m-1)N_X+1}\delta_{q,(m-1)N_X+1} \text{ (FEM)}, \tag{17}$$

$$\Sigma_2[p,q] = -jk_{m,N_X}at_{m,N_X}\delta_{p,mN_X}\delta_{q,mN_X}, \text{ (}m=1,2,...,M \text{ and } p,q=1,2,...,MN_X\text{) (FEM).} \tag{18}$$

By inserting Eqs. (13)-(14) or (15)-(18) into Eq. (12), we can evaluate the retarded Green's function, $G(E)$, at a given energy $E$. Then the spectral density functions due to the source/drain contacts can be obtained as[14]

$$A_1(E) = G(E)\Gamma_1(E)G^\dagger(E) \text{ and } A_2(E) = G(E)\Gamma_2(E)G^\dagger(E), \tag{19}$$

where $\Gamma_1(E) \equiv j\left(\Sigma_1(E) - \Sigma_1^\dagger(E)\right)$ and $\Gamma_2(E) \equiv j\left(\Sigma_2(E) - \Sigma_2^\dagger(E)\right)$, which determine the electron exchange rates between the active device region and the source/drain reservoirs at energy $E$. In this coupled mode space, the diagonal elements of the spectral function matrices represent the local density of states (LDOS) in the device for each mode. We define the LDOS for mode $m$ as $D_1^m$ (due to the source) and $D_2^m$ (due to the drain). Here $D_1^m$ and $D_2^m$ are both $N_X \times 1$ vectors obtained as

$$D_1^m [p] = \frac{1}{\pi a} A_1 \big[ (m-1) N_X + p, (m-1) N_X + p \big], \, (p=1,2,...,N_X) \qquad (20)$$

$$D_2^m [p] = \frac{1}{\pi a} A_2 \big[ (m-1) N_X + p, (m-1) N_X + p \big], \, (p=1,2,...,N_X). \qquad (21)$$

Then the 1D electron density (in m⁻¹) for mode $m$ can be calculated by

$$n_{1D}^m = \int_{-\infty}^{+\infty} \big[ D_1^m f \left( \mu_S, E \right) + D_2^m f \left( \mu_D, E \right) \big] dE , \qquad (22)$$

where $f$ is the Fermi-Dirac statistics function, and $\mu_S$ ($\mu_D$) is the source (drain) Fermi level, which is determined by the applied bias. The electron density obtained by Eq. (22) is a 1D distribution (along the x direction). To obtain a 3D electron density, we need couple Eq. (22) with the quantum confinement wavefunction for mode $m$,

$$n_{3D}^m \left( x, y, z \right) = n_{1D}^m \left( x \right) \big| \xi^m \left( y, z; x \right) \big|^2 . \qquad (23)$$

The total 3D electron density needs to be evaluated by summing the contributions from all the subbands in each conduction band valley. Then this 3D electron density is fed back to the Poisson solver for the self-consistent calculations. Once self-consistency is achieved, the electron current is computed by

$$I_{SD} = \frac{q}{\pi \hbar} \int_{-\infty}^{+\infty} T \left( E \right) \big[ f \left( \mu_S, E \right) - f \left( \mu_D, E \right) \big] dE , \qquad (24)$$

where the transmission coefficient, $T(E)$, can be evaluated as[14]

$$T \left( E \right) = trace \big[ \Gamma_1 \left( E \right) G \left( E \right) \Gamma_2 \left( E \right) G^\dagger \left( E \right) \big] . \qquad (25)$$

To obtain the total electron current, we also need to add up current components in all the conduction band valleys.

## B. The uncoupled mode space (UMS) approach

In the simulation of SNWTs, we assume that the shape of the Si body is uniform along the x direction. As a result, the confinement potential profile (in the y-z plane) varies very slowly along the channel direction. For instance, the conduction band-edge $U \left( x, y, z \right)$ takes the same shape but different values at different $x$. For this reason, the eigenfunctions $\xi^m \left( y, z; x \right)$ are approximately the same along the channel although the eigenvalues $E_{sub}^m(x)$ is different. So we assume

$$\xi^m(y,z;x) = \overline{\xi^m}(y,z) \tag{26}$$

or $\qquad \dfrac{\partial}{\partial x}\xi^m(y,z;x) = 0, \ (m{=}1,2,...,M),$ (27)

which infers

$$a_{mm}(x) = \overline{a_{mm}} = \oint_{y,z} \frac{1}{m_x^*(y,z)}\left|\overline{\xi^m}(y,z)\right|^2 dydz \ , \tag{28a}$$

$$b_{mn}(x) = 0 \ \text{ and } \ c_{mn}(x) = 0, \ (m,n{=}1,2,...,M). \tag{28b}$$

Inserting Eq. (28b) into Eq. (11), we obtain $h_{mn}=0$ ($m \neq n$ and $m,n{=}1,2,...,M$), which means that the coupling between the modes is negligible (all the modes are uncoupled). Thus the device Hamiltonian $H$ becomes a block-diagonal matrix

$$H = \begin{bmatrix} h_{11} & 0 & \cdots & \cdots & 0 \\ 0 & h_{22} & 0 & \ddots & \vdots \\ \vdots & 0 & \ddots & \ddots & \vdots \\ \vdots & \ddots & \ddots & \ddots & 0 \\ 0 & \cdots & \cdots & 0 & h_{MM} \end{bmatrix}. \tag{29}$$

Since all the input matrices at the RHS of Eq. (12) are either diagonal or block-diagonal, the retarded Green's function $G(E)$ is block-diagonal,

$$G(E) = \begin{bmatrix} G^1(E) & 0 & \cdots & \cdots & 0 \\ 0 & G^2(E) & 0 & \ddots & \vdots \\ \vdots & 0 & \ddots & \ddots & \vdots \\ \vdots & \ddots & \ddots & \ddots & 0 \\ 0 & \cdots & \cdots & 0 & G^M(E) \end{bmatrix}, \tag{30}$$

where $G^m(E)$ ($m{=}1,2,...,M$) is the Green's function for mode $m$ and is obtained as

$$G^m(E) = \left[ ES^m - h_{mm} - \Sigma_S^m(E) - \Sigma_1^m(E) - \Sigma_2^m(E) \right]^{-1}, \tag{31}$$

here $S^m$, $\Sigma_S^m$, $\Sigma_1^m$ and $\Sigma_2^m$ are all $N_X \times N_X$ matrices and defined as

$$S^m[p,q] = S\left[(m-1)N_X + p, (m-1)N_X + q\right], \ (p,q{=}1,2,...,N_X), \tag{32}$$

$$\Sigma_S^m[p,q] = \Sigma_S\left[(m-1)N_X + p, (m-1)N_X + q\right], \ (p,q=1,2,...,N_X), \quad (33)$$

$$\Sigma_1^m[p,q] = \Sigma_1\left[(m-1)N_X + p, (m-1)N_X + q\right], \ (p,q=1,2,...,N_X), \quad (34)$$

and

$$\Sigma_2^m[p,q] = \Sigma_2\left[(m-1)N_X + p, (m-1)N_X + q\right], \ (p,q=1,2,...,N_X). \quad (35)$$

Knowing the retarded Green's function, the spectral density functions due to the source/drain contacts for each mode $m$ can be obtained as[14]

$$A_1^m(E) = G^m(E)\Gamma_1^m(E)G^{m\dagger}(E) \text{ and } A_2^m(E) = G^m(E)\Gamma_2^m(E)G^{m\dagger}(E), \quad (36)$$

where $\Gamma_1^m(E) \equiv i\left(\Sigma_1^m(E) - \Sigma_1^{m\dagger}(E)\right)$ and $\Gamma_2^m(E) \equiv i\left(\Sigma_2^m(E) - \Sigma_2^{m\dagger}(E)\right)$. The LDOS for mode $m$, $D_1^m$ (due to the source) and $D_2^m$ (due to the drain), can then be evaluated by

$$D_1^m[p] = \frac{1}{\pi a}A_1^m[p,p] \text{ and } D_2^m[p] = \frac{1}{\pi a}A_2^m[p,p], \ (p=1,2,...,N_X). \quad (37)$$

After that, the electron charge density is computed by Eqs. (22) and (23). For the calculation of electron current, the total transmission coefficient can be written as a summation of the transmission coefficient $T^m(E)$ for each mode $m$,

$$T(E) = \sum_{m=1}^M T^m(E), \quad (38)$$

where $T^m(E)$ is obtained as[14]

$$T^m(E) = trace\left[\Gamma_1^m(E)G^m(E)\Gamma_2^m(E)G^{m\dagger}(E)\right]. \quad (39)$$

Finally, Eq. (38) is inserted into Eq. (24) to compute the electron current for the SNWT.

As we will show in Sec. III, for SNWTs, this uncoupled mode space approach shows excellent agreement with the CMS approach while maintaining higher computational efficiency. (The validity of the UMS approach for planar MOSFET simulation has been established by Venugopal et al[11] by doing a careful study of the UMS approach vs. 2D real space approach.)

## C. A fast uncoupled mode space (FUMS) approach

As described earlier, for both CMS and UMS approaches, we need to solve $N_X$ 2D Schrödinger equations (see Eq. (4)) in a self-consistent loop to obtain the electron subbands and eigenfunctions. For the device structures simulated in this work, this part of simulation usually takes more than 90% of the computational complexity, which makes parallel programming necessary. To increase the efficiency of our simulator and to make it executable on a single processor, we introduce a fast uncoupled mode space (FUMS) approach,[9,10] which only involves one 2D Schrödinger equation problem in a self-consistent loop and still provides excellent computational accuracy as compared with the CMS and UMS approaches. (The transport part of calculation in FUMS is the same as that in UMS.)

Recall the assumption made in part B that the eigenfunctions $\xi^m(y,z;x)$ are invariant along the x direction, $\xi^m(y,z;x) = \overline{\xi^m}(y,z)$ (Eq. (26)). Now we suppose that the average wavefunctions $\overline{\xi^m}(y,z)$ are the eigenfunctions of the following 2D Schrödinger equation

$$\left[ -\frac{\hbar^2}{2}\frac{\partial}{\partial y}\left(\frac{1}{m_y^*(y,z)}\frac{\partial}{\partial y}\right) - \frac{\hbar^2}{2}\frac{\partial}{\partial z}\left(\frac{1}{m_z^*(y,z)}\frac{\partial}{\partial z}\right) + \overline{U}(y,z) \right]\overline{\xi^m}(y,z) = \overline{E_{sub}^m} \cdot \overline{\xi^m}(y,z). \qquad (40)$$

Here the average conduction band-edge $\overline{U}(y,z)$ is obtained as

$$\overline{U}(y,z) = \frac{1}{L_X}\int_0^{L_X} U(x,y,z)\,dx, \qquad (41)$$

where $L_X$ is the total length of the simulated SNWT (including the S/D extensions). After computing the eigenvalues $\overline{E_{sub}^m}$ and eigenfunctions $\overline{\xi^m}(y,z)$ of this Schrödinger equation, we use the first order stationery perturbation theory to obtain the subband profile as[9,10]

$$E_{sub}^m(x) = \overline{E_{sub}^m} + \oint_{y,z} U(x,y,z)\left|\overline{\xi^m}(y,z)\right|^2 dydz - \oint_{y,z}\overline{U}(y,z)\left|\overline{\xi^m}(y,z)\right|^2 dydz. \qquad (42)$$

So far the subbands $E_{sub}^m(x)$ and the corresponding eigenfunctions $\xi^m(y,z;x)$ have been obtained approximately by only solving one 2D Schrödinger equation. The simulation results in Sec. III show that this FUMS approach has great accuracy for the calculation of both internal characteristics (e.g., the subband profiles) and terminal currents. The use of the FUMS approach highly improves the efficiency of our simulator and makes it a practical model for extensive device simulation and design.[18] (The simulation of a ballistic SNWT with 10nm gate length and 3nm Si body thickness normally takes <15 minutes per bias point on one 1.2GHz ATHLON processor).

# III. RESULTS FOR BALLISTIC SILICON NANOWIRE TRANSISTORS

In this section, we first verify the validity of the FUMS approach by comparing its results with those obtained by the UMS and CMS approaches. Then we adopt the FUMS as a simulation tool to explore device physics (i.e., both internal characteristics and terminal currents) of ballistic Si nanowire transistors with different types of cross-sections (e.g., triangular, rectangular and cylindrical).

## A. Benchmarking of the FUMS approach

As mentioned in Sec. II, for both CMS and UMS approaches, we need to solve a 2D Schrödinger equation (shown in Eq. (4)) at *each* slice of the SNWT to obtain the electron subbands and the corresponding eigenfunctions (modes). Figure 2 shows the electron wavefunctions at a slice of the SNWTs with a triangular, rectangular or cylindrical cross-section, respectively. After solving all the $N_X$ 2D Schrödinger equations, the electron subband levels are obtained (see FIG. 3, circles). For the FUMS approach, however, only one 2D Schrödinger equation needs to be solved, and the subband profile can then be calculated by Eq. (42). Figure 3 clearly illustrates that this approximation method (solid lines) provides excellent agreement with the rigorous calculation (circles), which shows that the FUMS approach correctly computes the electron subbands in SNWTs.

Figure 4 compares the computed $I_{DS}$ vs. $V_{GS}$ characteristics for the simulated cylindrical SNWT by the FUMS (dashed lines), UMS (circles) and CMS (crosses) approaches, respectively. It is clear that all the three approaches are in excellent agreement (<0.5% error), thus indicating that the FUMS approach, which has much higher computational efficiency than CMS and UMS, is an attractive simulation tool for modeling Si nanowire transistors. Although the sample device structure we use in FIG. 3 and FIG. 4 is a cylindrical SNWT, our conclusion is also applicable for SNWTs with arbitrary cross-sections. In the following parts of this work, we will use the FUMS approach to investigate the device physics in various SNWTs.

## B. Device physics and characteristics

The NEGF transport model we use in this work provides an opportunity to illustrate the local density of states (LDOS) of the simulated SNWTs. Figure 5 shows the LDOS together with the electron subbands for a ballistic cylindrical SNWT with 10nm gate length and 3nm Si body thickness. Strong oscillations in the LDOS plot are clearly observed, which is due to the quantum mechanical reflection. To be specific, the states injected from the drain are reflected off the drain-to-source barrier at the high drain bias and these reflected states strongly interfere with the injected ones. At the source end, the states injected at energies around the source barrier are also reflected and interfere. It should be noted that the occurrence of quantum inference in ballistic SNWTs relies on the quantum coherence (complete preservation of electron phase information) inside the devices. If scattering (dephasing mechanism) is included, as we will see in Sec. IV, the quantum interference and the oscillations in the LDOS are smeared out. In addition, the presence of states below the first electron subband is also visible in the LDOS plot, which is caused by source-to-drain tunneling.[20]

Figure 6 plots the 1D electron density (in $m^{-1}$) profile along the channel of the simulated cylindrical SNWT. It is clearly observed that the oscillations in the LDOS of the device result in an oscillation in the 1D electron density, even at the room temperature and more apparent at low temperature (77K). In general, such an oscillation in the electron density profile occurs in all kinds of transistors with 1D channels (e.g., the carbon nanotube transistor[21]). It is interesting to mention that there is no evident oscillation in the electron density profile in a planar MOSFET (see FIG. 8 on p. 3736 in Ref. [11]) although its LDOS also bears strong oscillations (see FIG. 4 on p. 3735 in Ref. [17]). The reason is that in a planar MOSFET there is a transverse direction (normal to both the Si/SiO$_2$ interfaces and the channel direction), in which the electron wavefunction is assumed to be a plane wave, thus resulting in numerous transverse modes in the device. These transverse modes wash out the oscillations in the LDOS and cause a smooth electron density profile. So the oscillation in the electron density profile is a special property of SNWT as compared with planar MOSFETs.

Figure 7 illustrates the calculated transmission coefficient (from Eqs. (38) and (39)) for the simulated cylindrical SNWT. When the total electron energy increases above the source end of the first subband, the electrons start to be injected into the channel, so the transmission coefficient begins to increase from zero. As the electron energy continues to go up, the second and third subbands (modes) become conductive successively, which results the step-like shape of the transmission coefficient curve. We also observe that the transmission coefficient is above zero even when the total electron energy is below the top of barrier of the first subband, which is the evidence of source-to-drain tunneling.

In FIG. 8, we compare the $I_{DS}$ vs. $V_{GS}$ characteristics for SNWTs with triangular, rectangular and cylindrical cross-sections. Two interesting phenomena are evidently visible: 1) the cylindrical wire (CW) and triangular wire (TW) transistors have higher threshold voltages, $V_{TH}$ (that is defined as $I_{DS}(V_{GS}=V_{TH})=10^{-8}A$ when $V_{DS}=0.4V$), due to stronger quantum confinement (the cross-section areas of the CW and TW are smaller than that of the RW for the same Si body thickness), and 2) the CW SNWT offers the best subthreshold swing and the highest on-off current ratio (under the same gate overdrive, $V_{GS}$-$V_{TH}$) due to its good gate control. These results clearly show that our simulator correctly treats the 3D electrostatics, quantum confinement and transport in SNWTs with arbitrary cross-sections.

## IV. TREATMENT OF SCATTERING WITH BÜTTIKER PROBES

In this section, we apply a simple quantum treatment of scattering based on the Büttiker probes[14,16,17] to our SNWT simulation. The simulation results show that this simple model captures the essential effects of scattering on both internal device parameters (e.g., charge distribution and electrostatic potential) and current-voltage characteristics. (A detailed treatment of scattering within the NEGF formalism is important to deeply understand physics in Si nanowires, and it will be discussed in future work.)

### A. Theory

The simple treatment of scattering with the Büttiker probes has been adopted by Venugopal and coworkers[17] for the simulation of nanoscale MOSFETs. Due to the similarity between the transport calculations of a MOSFET and an SNWT, here we will follow the basic concepts and formalism of the method described in Ref. [17] while making necessary modifications and corrections for the case of SNWT simulation.

In the ballistic regime, as we know, electrons move through the device coherently, with their energies and phase information conserved. When scattering is present, however, electrons' momenta and energies could be altered and their phase information may be lost. Based on this observation, virtual probes (Büttiker probes) are attached to the device lattice (in the channel direction), which serve as reservoirs that absorb electrons from the active device, modulate their momenta and/or energies, and then reinject them back to the device. The difference between the probes and the S/D contacts is that the probes can only change the electron momentum/energy and not the number of electrons within the active device.[17]

Figure 9 shows the 1D device lattice (in the channel direction) for an SNWT with the Büttiker probes attached. Each probe is treated as a virtual 1D lattice (in the $x'$ direction) that is coupled to a node in the device lattice. The coupling energy, $\Delta_m^i$, between this virtual lattice and the node it is attached to is called the Büttiker probe strength,[17] which is determined by the ballisticity of the device. For instance, when $\Delta_m^i$ is zero, there is no coupling between the device and the probes, so the electrons can travel through the device ballistically. If this coupling energy is large, it means that the electrons in the active device region can easily scatter into the probes, which implies that the scatting in the device is strong. As we will show later, the Büttiker probe strength can be analytically related to the electron mean-free-path, which allows us to calibrate the parameters in our simulation to mimic a low field mobility that can be measured experimentally.[17] It should also be noted that since we treat each probe as a reservoir, a Fermi level ($\mu_i$, $i = 2,...,N_X - 1$) needs to be assigned to the probe, and the values of these probe Fermi levels have to be adjusted to achieve current continuity (i.e., the net cuurent at each probe is zero). The mathematical formalism used to treatment this physical structure is described in the following paragraphs.

As we show in Sec. II, the retarded Green's function for mode $m$ is obtained as

$$G^m(E) = \left[ ES^m - h_{mm} - \Sigma_S^m(E) - \Sigma_1^m(E) - \Sigma_2^m(E) \right]^{-1}. \tag{31}$$

If we discretize the matrices by the FDM method, $S^m$ is a $N_X \times N_X$ identity matrix and the device Hamiltonian $h_{mm}$ is expressed as

$$
h_{mm} = \begin{bmatrix}
2t_m + E_{sub}^m(0) & -t_m & 0 & \cdots & \cdots & 0 \\
-t_m & 2t_m + E_{sub}^m(a) & -t_m & \ddots & & \vdots \\
0 & -t_m & \ddots & \ddots & & \vdots \\
\vdots & \ddots & \ddots & \ddots & -t_m & 0 \\
\vdots & & \ddots & -t_m & 2t_m + E_{sub}^m\big((N_X-2)a\big) & -t_m \\
0 & \cdots & \cdots & 0 & -t_m & 2t_m + E_{sub}^m\big((N_X-1)a\big)
\end{bmatrix} \quad \text{(FDM), (43)}
$$

where the coupling energy between adjacent lattice nodes (in the $x$ direction) is $t_m = \left(\hbar^2 / 2a^2\right)\overline{a_{mm}}$ and $\overline{a_{mm}}$ is defined in Eq. (28a). In the *ballistic* limit, the scattering self-energy $\Sigma_S^m = 0$ so the total self-energy matrix is

$$
\Sigma^m = \Sigma_S^m + \Sigma_1^m + \Sigma_2^m = \begin{bmatrix}
-t_m e^{ik_{m,1}a} & 0 & \cdots & \cdots & 0 \\
0 & 0 & \ddots & \ddots & \vdots \\
\vdots & \ddots & \ddots & \ddots & \vdots \\
\vdots & \ddots & \ddots & 0 & 0 \\
0 & \cdots & \cdots & 0 & -t_m e^{ik_{m,N_X}a}
\end{bmatrix} \quad \text{(FDM),} \qquad (44)
$$

where $k_{m,1}$ ( $k_{m,N_X}$ ) is determined by $E = E_{sub}^m(0) + 2t_m\big(1 - \cos k_{m,1}a\big)$ ( $E = E_{sub}^m\big((N_X-1)a\big) + 2t_m\big(1 - \cos k_{m,N_X}a\big)$ ). After we attach the Büttiker probes to the device lattice (FIG. 9), the device Hamiltonian $h_{mm}$ becomes

$$
h_{mm} = \begin{bmatrix}
2t_m + E_{sub}^m(0) & -t_m & 0 & \cdots & \cdots & 0 \\
-t_m & 2t_m + \Delta_m^2 + E_{sub}^m(a) & -t_m & \ddots & & \vdots \\
0 & -t_m & \ddots & \ddots & & \vdots \\
\vdots & \ddots & \ddots & \ddots & -t_m & 0 \\
\vdots & & \ddots & -t_m & 2t_m + \Delta_m^{N_X-1} + E_{sub}^m\big((N_X-2)a\big) & -t_m \\
0 & \cdots & \cdots & 0 & -t_m & 2t_m + E_{sub}^m\big((N_X-1)a\big)
\end{bmatrix}, (45)
$$

and the total self-energy matrix turns to

$$
\Sigma^m = \Sigma_S^m + \Sigma_1^m + \Sigma_2^m = \begin{bmatrix}
-t_m e^{ik_{m,1}a} & 0 & \cdots & \cdots & 0 \\
0 & -\Delta_m^2 e^{ik_{m,2}a} & 0 & \ddots & \vdots \\
\vdots & \ddots & \ddots & \ddots & \vdots \\
\vdots & \ddots & 0 & -\Delta_m^{N_X-1} e^{ik_{m,N_X-1}a} & 0 \\
0 & \cdots & \cdots & 0 & -t_m e^{ik_{m,N_X}a}
\end{bmatrix}, \qquad (46)
$$

where $k_{m,i}$ $(i=1,2,...,N_X)$ is determined by $E = E_{sub}^{m}\left((i-1)a\right)+2t_m\left(1-\cos k_{m,i}a\right)$, and $\Delta_m^i$ $(i=2,3,...,N_X-1)$ is the Büttiker probe strength. For convenience, we prefer to keep the device Hamiltonian $h_{mm}$ in its original form (Eq. (43)), so we have to move the terms containing $\Delta_m^i$ in the diagonal elements of $h_{mm}$ to the total self-energy matrix $\Sigma^m$. Thus,

$$\Sigma^m = \Sigma_S^m + \Sigma_1^m + \Sigma_2^m = \begin{bmatrix} -t_m e^{ik_{m,1}a} & 0 & \cdots & & \cdots & 0 \\ 0 & -\Delta_m^2\left(e^{ik_{m,2}a}-1\right) & 0 & & \ddots & \vdots \\ \vdots & \ddots & \ddots & & \ddots & \vdots \\ \vdots & \ddots & & 0 & -\Delta_m^{N_X-1}\left(e^{ik_{m,N_X-1}a}-1\right) & 0 \\ 0 & \cdots & & \cdots & 0 & -t_m e^{ik_{m,N_X}a} \end{bmatrix}. \quad (47)$$

Inserting Eqs.(43) and (47) into Eq. (31), the retarded Green's function $G^m$ can be evaluated.

Knowing $G^m$, the state spectral function due to injection from the S/D and all probes for mode $m$ is obtained as[17]

$$A_i^m\left(E\right) = G^m\left(E\right)\Gamma_i^m\left(E\right)G^{m\dagger}\left(E\right), \quad (48)$$

where $i$ runs over all the reservoirs (including the S/D) and $\Gamma_i^m$ is an $N_X \times N_X$ matrix defined as

$$\Gamma_i^m\left[p,q\right] = j\left[\Sigma^m\left[p,q\right]-\Sigma^{m\dagger}\left[p,q\right]\right]\delta_{p,i}\delta_{q,i}, \ (p,q=1,2,...,N_X). \quad (49)$$

The local density of states due to injection from reservoir $i$ is then obtained as

$$D_i^m\left[p\right] = \frac{1}{\pi a}A_i^m\left[p,p\right], \ (i=1,2,...,N_X, \ p=1,2,...,N_X), \quad (50)$$

and the 1D electron density (in m$^{-1}$) for mode $m$ can be calculated by

$$n_{1D}^m = \sum_i \int_{-\infty}^{+\infty} D_i^m f\left(\mu_i,E\right)dE, \quad (51)$$

where $i$ is the reservoir index that runs over all the probes and the S/D, and $\mu_i$ is the Fermi level for reservoir $i$ (note that $\mu_1 = \mu_S$ and $\mu_{N_X} = \mu_D$).

The transmission coefficient between any two reservoirs $i$ and $r$ can be evaluated as

$$T_{i\leftrightarrow r}^m\left(E\right) = trace\left[\Gamma_i^m\left(E\right)G^m\left(E\right)\Gamma_r^m\left(E\right)G^{m\dagger}\left(E\right)\right]. \quad (52)$$

The net current density (at energy $E$) at reservoir $i$ including contributions from all reservoirs (labeled by $r$), modes (labeled by $m$) and valleys is

$$\eta_i(E) = \frac{q}{\pi\hbar} \sum_m \sum_r T_{i\leftrightarrow r}^m(E) \left[ f(\mu_i, E) - f(\mu_r, E) \right],$$ (53)

and the net current at reservoir $i$ is

$$I_i = \int_{-\infty}^{+\infty} \eta_i(E) dE.$$ (54)

As mentioned in Ref. [17], while the S/D Fermi levels are determined by the applied voltages, the Fermi levels of the probes have to be adjusted to ensure current continuity, which implies that the net current at each probe must be zero,

$$I_i = \int_{-\infty}^{+\infty} \eta_i(E) dE = 0, \ (i = 2, 3, ..., N_X - 1).$$ (55)

Inserting Eq. (53) into (55), we obtain

$$\frac{q}{\pi\hbar} \sum_m \sum_r \int_{-\infty}^{+\infty} T_{i\leftrightarrow r}^m(E) \left[ f(\mu_i, E) - f(\mu_r, E) \right] dE = 0, \ (i = 2, 3, ..., N_X - 1).$$ (56)

Solving this nonlinear equation group (56) by Newton's method,[17] the Fermi levels ($\mu_i$, $i = 2, 3, ..., N_X - 1$) of all the probes are evaluated. It should be mentioned that if we implement the elastic Büttiker probes, which can only change the electron momentum and not the energy, to capture elastic scattering mechanisms in SNWTs (e.g., surface roughness scattering and ionized impurity scattering), the net current for each probes has to be zero at any energy, so

$$\eta_i(E) = \frac{q}{\pi\hbar} \sum_m \sum_r T_{i\leftrightarrow r}^m(E) \left[ f(\mu_i, E) - f(\mu_r, E) \right] = 0, \ (i = 2, 3, ..., N_X - 1).$$ (57)

It implies that the probe Fermi levels are both position and energy dependent. In this case, the Fermi levels of probes at each energy can be computed by solving the linear equation group (57). Knowing the probe Fermi levels (by solving either Eq. (56) or Eq. (57)), the electron density and terminal current can be calculated from Eqs. (51) and (54).

Finally, we list the equations that relate the Büttiker probe strength, $\Delta_m^i$, to the classical low field electron mobility $\mu_0$. Following the procedures in Ref. [17], for a single mode 1D conductor with a uniform potential, we can obtain

$$\frac{\Delta_m^i}{t_m} = \frac{2a}{\lambda},$$ (58)

where $\lambda$ is the electron mean-free-path, which relates to the low field electron mobility by the following equation for a 1D conductor (the $\lambda \sim \mu_0$ relation for a 2D conductor is described in Ref. [23]),

$$\lambda = \left( \frac{2\mu_0}{\upsilon_T} \frac{k_B T}{q} \right) \cdot \frac{[\Im_{-1/2}(\eta_F^i)]^2}{\Im_{-3/2}(\eta_F^i)\Im_0(\eta_F^i)}, \tag{59}$$

where $\upsilon_T = \sqrt{2k_B T / \pi m_x^*}$ is the uni-directional thermal velocity of non-degenerate electrons. The function, $\Im_n(x)$, is the Fermi-Dirac integral, and $\eta_F^i$ is defined as $\eta_F^i = (\mu_i - E_{sub}^m(x_i))/k_B T$, where $x_i$ is the position of the $i^{th}$ reservoir (probe) of the device. It should be noted that the mean-free-path $\lambda$ defined in Eq. (59) is position-dependent and consequently the Büttiker probe strength, $\Delta_m^i$, is also position-dependent. As mentioned earlier, single mode occupancy is assumed in our analysis. If more than one mode is occupied, the mean-free-path should be treated as an average mean-free-path over all the modes and valleys. (Please refer to Appendix B in Ref. [17] for details.)

## B. Results

Figure 10 plots the LDOS together with the electron subbands for a dissipative cylindrical SNWT with 10nm gate length and 3nm Si body thickness. We assume that both elastic (e.g., surface roughness scattering and ionized impurity scattering) and inelastic (e.g., electron-phonon interactions) scattering mechanisms are present in the device (i.e., Eq. (56) is used for current continuity), and the equivalent mobility is $55 cm^2 /(V \cdot s)$ at the S/D extension regions and $200 cm^2 /(V \cdot s)$ in the channel. Compared with the ballistic case (FIG. 5), strong oscillations in the LDOS, which is due to quantum interference, are washed out. It is because scattering inside the SNWT randomizes the phase of the electrons and consequently destroys the quantum coherence in the device.[14,17] Moreover, the slope of the electron subbands in the S/D extension regions manifests the S/D series resistances at the on-state, which is caused by the strong scattering (i.e., the S/D mobility is only $55 cm^2 /(V \cdot s)$) at the heavily doped S/D regions. In FIG. 11, we compare the $I_{DS}$ vs. $V_{GS}$ characteristics for this dissipative cylindrical SNWT (solid lines) with its ballistic limit (dashed lines). It is evidently shown that scattering lowers both off and on currents. For the mobility values we use in the simulation, the on-current of the dissipative SNWT approaches ~70% of the ballistic limit.

The above results clearly indicate that the simple quantum treatment of scattering with the Büttiker probes captures the effects of scattering on both internal characteristics and terminal currents for SNWTs. The relation between the Büttiker probe strength, the only input parameter in this model, with the experimentally measurable low field mobility enables this simple model to be used in engineering simulation and design. It should also be noted, however, that this phenomenological model is only a macroscopic description of scattering, which is similar as the drift-diffusion model that is used in the semiclassical context. To quantum mechanically treat various scattering mechanisms in detail, a rigorous quantum treatment of scattering within the NEGF formalism[14] is still needed.

## V. SUMMARY

In this paper, we present a computationally efficient three-dimensional quantum simulation of various silicon nanowire transistors based on the effective-mass approximation. The coupled/uncoupled mode space approaches are adopted to decompose the 3D device Hamiltonian, which greatly reduces the simulation time while keeping excellent computational accuracy. The use of a fast uncoupled mode space further scales down the computational complexity and makes our simulator executable on a single processor. This enables our approach to be used as a practical 3D quantum model for extensive device simulation and design.

Although we mainly focus on ballistic simulations in this work, a simple treatment of scattering with the Büttiker probes, previously applied to MOSFET simulations, is also implemented in our SNWT simulator. This model is a one input parameter model and the parameter we use can be related to the experimentally measurable low field mobility of electrons. As a result, the implementation of this simple scattering model endows our SNWT simulator with the ability to explore the realistic performance limits of SNWTs.

# ACKNOWLEDGEMENTS

This work was supported by the Semiconductor Research Corporation (SRC) and the NSF Network for Computational Nanotechnology (NCN). The authors would like to thank Prof. Supriyo Datta, Anisur Rahman, Dr. Avik Ghosh, Dr. Ramesh Venugopal (currently at Texas Instruments), Jing Guo and other group members for their sincere help.

## Figure Captions:

FIG. 1 The simulated SNWT structures in this work. (a) A schematic graph of an intrinsic SNWT with arbitrary cross-sections (for clarity, the SiO2 substrate is not shown here). (b) The gird used in the simulation of SNWTs. (c) The cross-sections of the simulated triangular wire (TW), rectangular wire (RW) and cylindrical wire (CW) FETs. $T_{Si}$ is the silicon body thickness, $W_{Si}$ is the silicon body width and $W_{Wire}$ is the wire width. For the TW, the direction normal to each gate is <111>, so the channel is <101> oriented. In contrast, for the channel of the RW, both <101> and <100> orientations are possible. For the CW, we assume the channel to be <100> oriented.

FIG. 2 The 2D Modes (the square of the modulus of the electron wavefunctions in the (010) valleys) in a slice of (a) triangular wire (TW), (b) rectangular wire (RW) and (c) cylindrical wire (CW) transistors. For clarity, the SiO$_2$ substrates for TW and RW FETs are not shown here.

FIG. 3 The electron subband profile in a cylindrical SNWT with 10nm gate length ($V_{GS}=0.4V$ and $V_{DS}=0.4V$). The silicon body thickness, $T_{Si}$ (as shown in FIG. 1 (c)), is 3nm, and the oxide thickness is 1nm. The source/drain (S/D) doping concentration is $2\times10^{20}$cm$^{-3}$ and the channel is undoped. The solid lines are for the approximation method (solving a 2D Schrödinger equation only once) used in the FUMS approach, while the circles are for the rigorous calculation (solving 2D Schrödinger equations $N_X$ times) adopted in the UMS and CMS approaches.

FIG. 4 The $I_{DS}$ vs. $V_{GS}$ curves for a cylindrical SNWT in logarithm (left) and linear (right) scales ($V_{DS}=0.4V$). The device structure is the same as that in FIG. 3. The crosses are for the CMS approach, the circles are for the UMS approach, and the dashed lines are for the FUMS approach.

FIG. 5 The computed LDOS (in $1/(eV \cdot m)$) and electron subbands (dashed lines) of a ballistic cylindrical SNWT with 10nm gate length and 3nm Si body thickness (the details of the device geometry are described in the FIG. 3 caption). ($V_{GS}$=0.4V and $V_{DS}$=0.4V).

FIG. 6 The 1D electron density profile along the channel of the simulated cylindrical SNWT (the details of the device geometry are described in the FIG. 3 caption). The solid line is for $T$=300K while the dashed line is for $T$=77K. ($V_{GS}$=0.4V and $V_{DS}$=0.4V).

FIG. 7 The transmission coefficient and electron subbands in the simulated cylindrical SNWT (the details of the device geometry are described in the FIG. 3 caption). ($V_{GS}$=0.4V and $V_{DS}$=0.4V).

FIG. 8 The $I_{DS}$ vs. $V_{GS}$ curves for the triangular wire (TW) FET with <101> oriented channels, rectangular wire (RW) FET with <101> oriented channels and cylindrical wire (CW) FET with <100> oriented channels. ($V_{DS}$=0.4V). All the SNWTs have the same silicon body thickness ($T_{Si}$=3nm), oxide thickness ($T_{ox}$=1nm), gate length ($L$=10nm) and gate work function ($WF$=4.05eV). The Si body width, $W_{Si}$, of the RW is 4nm. In the calculation of the TW and RW FETs, whose channels are <101> oriented, the effective masses of electrons in the (100) and (001) valleys are obtained from Ref. [22] as $m_x^* = 0.585m_e$, $m_y^* = 0.19m_e$ and $m_z^* = 0.318m_e$.

FIG. 9 A generic plot of the 1D device lattice (solid line with dots, along the $X$ direction) for an SNWT with the Büttiker probes attached. Each probe is treated as a virtual 1D lattice (dashed line with dots, along the $X'$ direction) that is coupled to a node in the device lattice. The coupling energy between this virtual lattice and the node it is attached to is $\Delta_m^i$, and that between two adjacent device lattice nodes is $t_m$. The probe Fermi levels are labeled as $\mu_i$ ($i = 2,3,...,N_X - 1$).

FIG. 10 The computed LDOS (in $1/(eV \cdot m)$) and electron subbands (dashed lines) of a dissipative cylindrical SNWT with 10nm gate length and 3nm Si body thickness (the details of the device geometry are described in the FIG. 3 caption). ($V_{GS}$=0.4V and $V_{DS}$=0.4V). The S/D mobility is $55cm^2/(V \cdot s)$ and the channel mobility is $200cm^2/(V \cdot s)$.

FIG. 11 The $I_{DS}$ vs. $V_{GS}$ curves for a cylindrical SNWT with 10nm gate length and 3nm Si body thickness (the details of the device geometry are described in the FIG. 3 caption) in logarithm (left) and linear (right) scales ($V_{DS}$=0.4V). The dashed lines are for the ballistic limit while the solid lines are for the case with scattering (i.e., the S/D mobility is $55cm^2/(V \cdot s)$ and the channel mobility is $200cm^2/(V \cdot s)$).

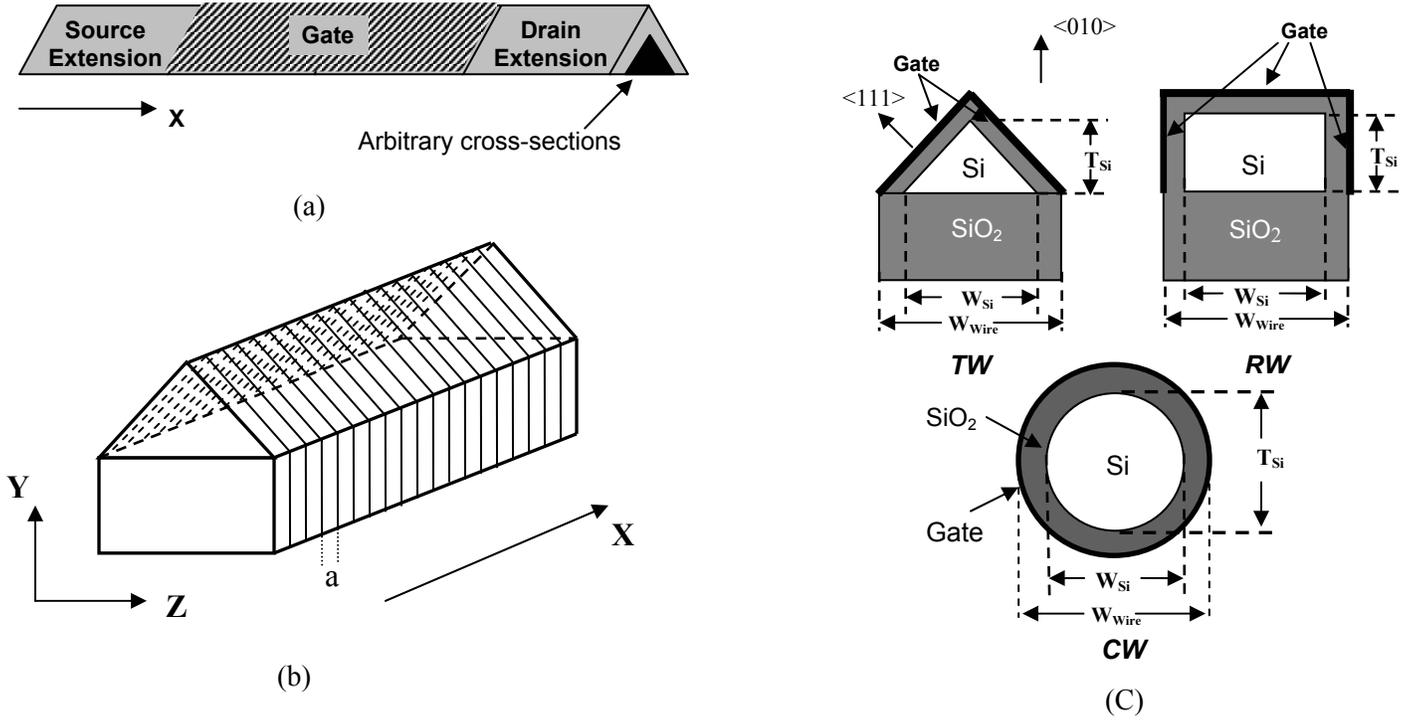

(a)

(b)

(C)

FIG. 1 The simulated SNWT structures in this work. (a) A schematic graph of an intrinsic SNWT with arbitrary cross-sections (for clarity, the SiO₂ substrate is not shown here). (b) The gird used in the simulation of SNWTs. (c) The cross-sections of the simulated triangular wire (TW), rectangular wire (RW) and cylindrical wire (CW) FETs. $T_{Si}$ is the silicon body thickness, $W_{Si}$ is the silicon body width and $W_{Wire}$ is the wire width. For the TW, the direction normal to each gate is <111>, so the channel is <101> oriented. In contrast, for the channel of the RW, both <101> and <100> orientations are possible. For the CW, we assume the channel to be <100> oriented.

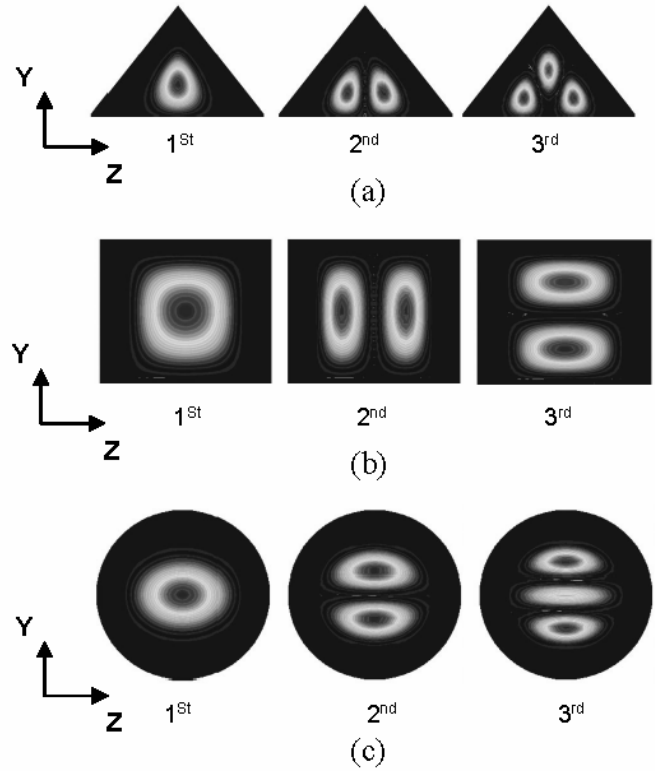

FIG. 2 The 2D Modes (the square of the modulus of the electron wavefunctions in the (010) valleys) in a slice of (a) triangular wire (TW), (b) rectangular wire (RW) and (c) cylindrical wire (CW) transistors. For clarity, the SiO$_2$ substrates for TW and RW FETs are not shown here.

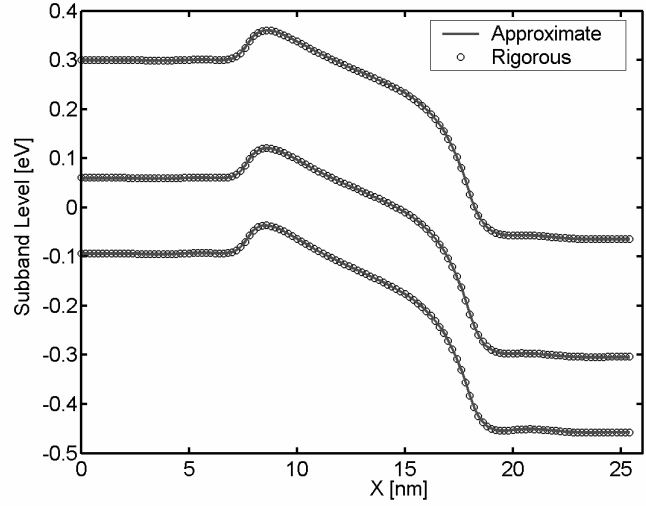

FIG. 3 The electron subband profile in a cylindrical SNWT with 10nm gate length ($V_{GS}$=0.4V and $V_{DS}$=0.4V). The silicon body thickness, $T_{Si}$ (as shown in FIG. 1 (c)), is 3nm, and the oxide thickness is 1nm. The source/drain (S/D) doping concentration is $2\times10^{20}$cm$^{-3}$ and the channel is undoped. The solid lines are for the approximation method (solving a 2D Schrödinger equation only once) used in the FUMS approach, while the circles are for the rigorous calculation (solving 2D Schrödinger equations $N_X$ times) adopted in the UMS and CMS approaches.

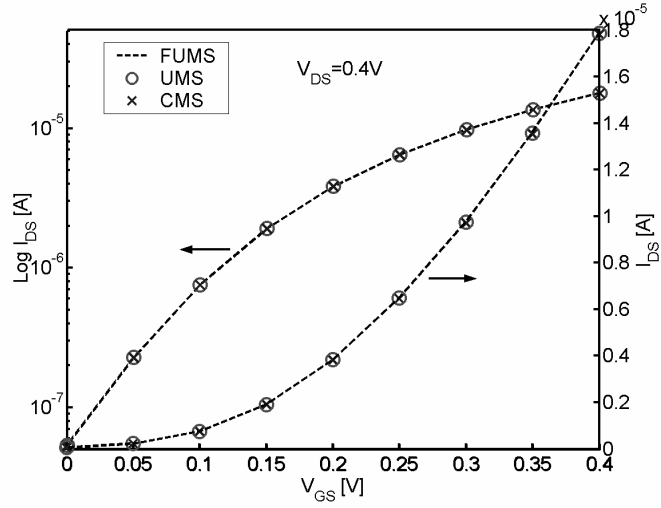

FIG. 4 The $I_{DS}$ vs. $V_{GS}$ curves for a cylindrical SNWT in logarithm (left) and linear (right) scales ($V_{DS}=0.4V$). The device structure is the same as that in FIG. 3. The crosses are for the CMS approach, the circles are for the UMS approach, and the dashed lines are for the FUMS approach.

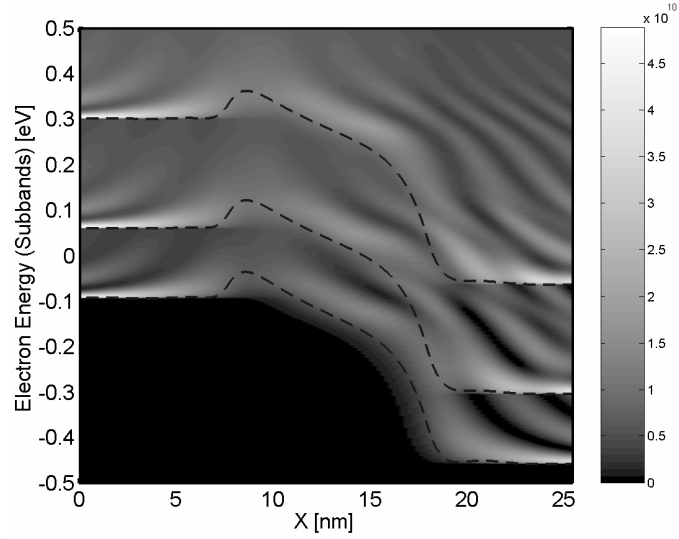

FIG. 5 The computed LDOS (in $1/(eV \cdot m)$) and electron subbands (dashed lines) of a ballistic
cylindrical SNWT with 10nm gate length and 3nm Si body thickness (the details of the
device geometry are described in the FIG. 3 caption). ($V_{GS}=0.4V$ and $V_{DS}=0.4V$).

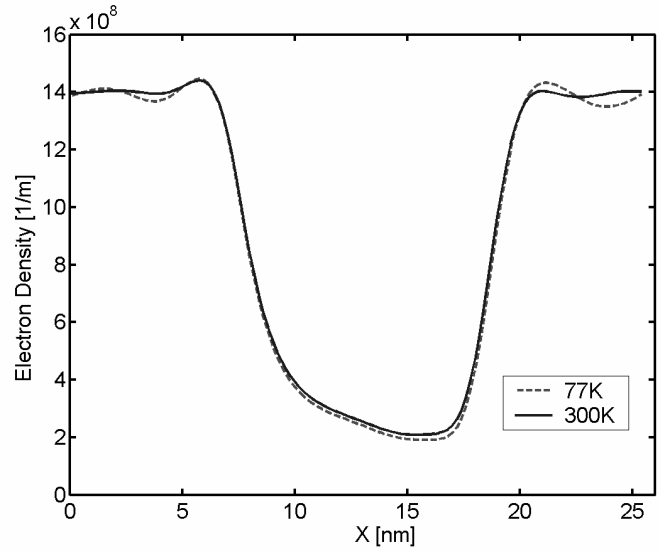

FIG. 6 The 1D electron density profile along the channel of the simulated cylindrical SNWT (the details of the device geometry are described in the FIG. 3 caption). The solid line is for $T=300K$ while the dashed line is for $T=77K$. ($V_{GS}=0.4V$ and $V_{DS}=0.4V$).

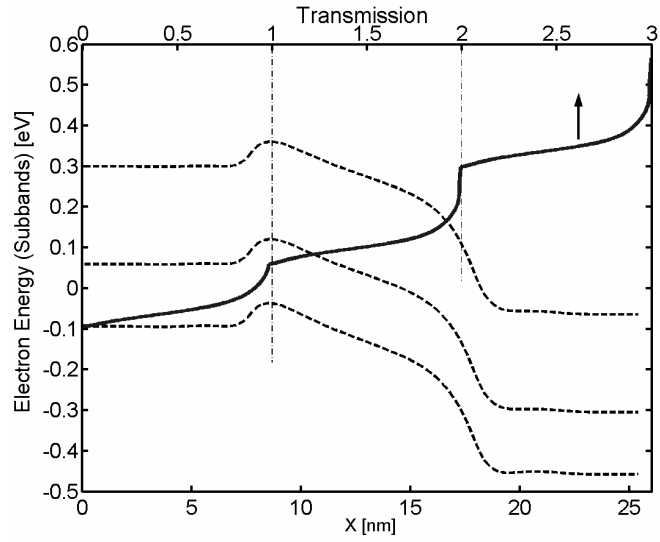

FIG. 7 The transmission coefficient and electron subbands in the simulated cylindrical SNWT (the details of the device geometry are described in the FIG. 3 caption). ($V_{GS}$=0.4V and $V_{DS}$=0.4V).

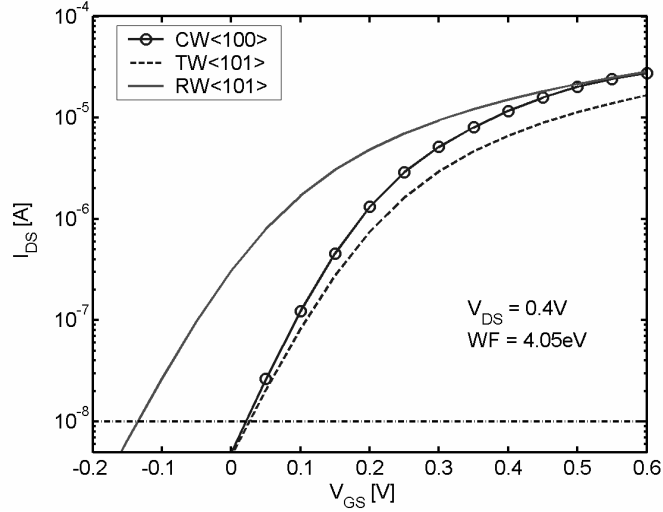

FIG. 8 The $I_{DS}$ vs. $V_{GS}$ curves for the triangular wire (TW) FET with <101> oriented channels, rectangular wire (RW) FET with <101> oriented channels and cylindrical wire (CW) FET with <100> oriented channels. ($V_{DS}=0.4V$). All the SNWTs have the same silicon body thickness ($T_{Si}=3nm$), oxide thickness ($T_{ox}=1nm$), gate length ($L=10nm$) and gate work function ($WF=4.05eV$). The Si body width, $W_{Si}$, of the RW is 4nm. In the calculation of the TW and RW FETs, whose channels are <101> oriented, the effective masses of electrons in the (100) and (001) valleys are obtained from Ref. [22] as $m_x^* = 0.585m_e$, $m_y^* = 0.19m_e$ and $m_z^* = 0.318m_e$.

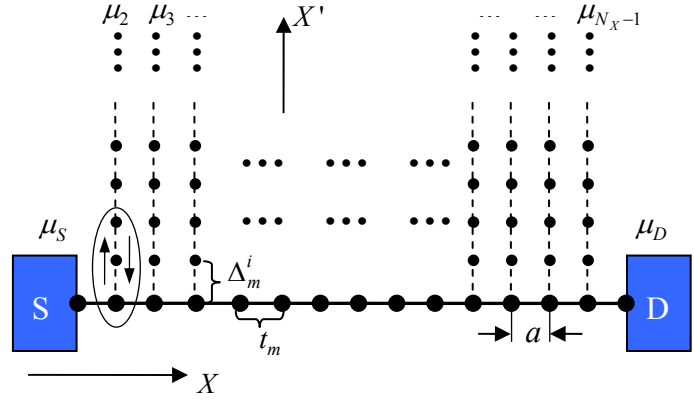

FIG. 9 A generic plot of the 1D device lattice (solid line with dots, along the $X$ direction) for an SNWT with the Büttiker probes attached. Each probe is treated as a virtual 1D lattice (dashed line with dots, along the $X'$ direction) that is coupled to a node in the device lattice. The coupling energy between this virtual lattice and the node it is attached to is $\Delta_m^i$, and that between two adjacent device lattice nodes is $t_m$. The probe Fermi levels are labeled as $\mu_i$ ($i = 2,3,...,N_X - 1$).

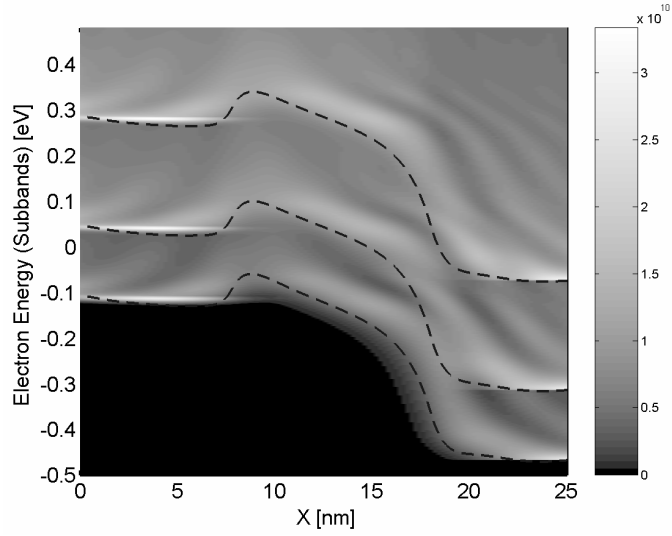

FIG. 10 The computed LDOS (in $1/(eV \cdot m)$) and electron subbands (dashed lines) of a dissipative cylindrical SNWT with 10nm gate length and 3nm Si body thickness (the details of the device geometry are described in the FIG. 3 caption). ($V_{GS}$=0.4V and $V_{DS}$=0.4V). The S/D mobility is $55cm^2/(V \cdot s)$ and the channel mobility is $200cm^2/(V \cdot s)$.

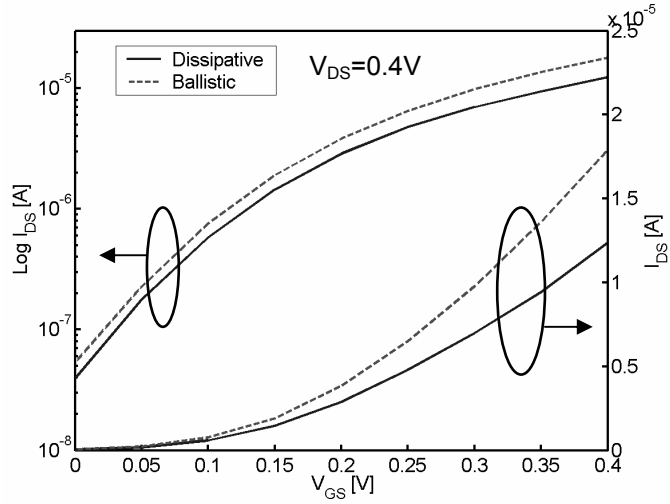

FIG. 11 The $I_{DS}$ vs. $V_{GS}$ curves for a cylindrical SNWT with 10nm gate length and 3nm Si body thickness (the details of the device geometry are described in the FIG. 3 caption) in logarithm (left) and linear (right) scales ($V_{DS}=0.4V$). The dashed lines are for the ballistic limit while the solid lines are for the case with scattering (i.e., the S/D mobility is $55 cm^2/(V \cdot s)$ and the channel mobility is $200 cm^2/(V \cdot s)$ ).